
\magnification=1200
\vsize=24 true cm
\hsize=15.2 true cm
\baselineskip=0.6 true cm

\parindent=1 true cm
\pageno=1

\def\lsim{\mathrel{\rlap{\lower4pt\hbox{\hskip1pt$\sim$}}
\raise1pt\hbox{$<$}}}         
\def\gsim{\mathrel{\rlap{\lower4pt\hbox{\hskip1pt$\sim$}}
\raise1pt\hbox{$>$}}}         

\vglue 1 true cm

\centerline{NEW MEASUREMENT OF THE SINGLE DIFFRACTION DISSOCIATION}
\centerline{AND THE NATURE OF THE POMERON}

\bigskip
\centerline{\bf L.L.Jenkovszky and B.V.Struminsky}
\bigskip
\centerline{Bogoliubov Insitut for Theoretical Physics,}
\centerline{Kiev-143, Ukraine;}
\centerline{E-mail: JENK@GLUK.APC.ORG}
\bigskip
\centerline{\bf Abstract}
\smallskip
Recent data on single diffraction dissotiation, measured by CDF at the
Tevatron show a moderate increase with energy of the cross section,
thus confirming the predictions based on the dipole
pomeron model.

 \bigskip

{\bf 1.Introduction}
\smallskip
Recently new results on single diffraction dissociation
from  the Tevatron at $\sqrt s=546\ (1800)\ GeV$  have been published
by the CDF [1].  The measured integrated cross section
$$\sigma_{SD}=7.89\pm0.33\ (9.46\pm0.44)\  mb,$$
was found to be much below the value expected from extrapolations
based on the popular supercritical pomeron model.

In this connection we find it appropriate to recall the
results of our earlier investigations based on the dipole pomeron
model [2]. In that paper, we have used a unitarized (so-called
$u$-matrix) representation for the scattering amplitude of the form
$$T(\rho,s)={{u(\rho,s)}\over{1-iu(\rho,s)}}, \eqno(1)$$
where $\rho$ is the impact parameter.

The $u$-matrix was choosen in the form
$$u(\rho,s)=ig(s)\exp\bigg(-{\rho^2\over{4\alpha'(b+L)}}\bigg),
\eqno(2)$$
coresponding to a dipole Pomeron in the impact parameter
repesentation [2] ($L=\ln(s/s_0),$ and $b$, $s_0$ are adjustable
parameters).
{}From eqs.  (1) and (2) one obtains [2] for the total
and integrated elastic scattering cross sections:
$$\sigma_t=16\pi\alpha'(b+L)\ln(1+g),\eqno (3) $$
$$\sigma_{el}=16\pi\alpha'(b+L)\big[\ln(1+g)-{g\over{1+g}}\big],\eqno (4) $$
$$\sigma_{in}=16\pi\alpha'(b+L){g\over{1-g}}.\eqno (5)$$

The rise with the energy of the ratio $\sigma_{el}/\sigma_{tot}$ has
motivated the choice of $g=g(s)$ in the form
$$g(s)=g_0\big(s/s_0)^{\epsilon},$$
where $\epsilon=0.06.$ It corresponds to a double Regge (Pomeron)
pole with the intercept $1+\epsilon.$  In this model, the total cross
sections at asymptotic energies saturate the Froisart bound,
$\ln^2(s),$ but at present accelerator energies the rate of increase
is between $\ln(s)$ and $\ln^2(s)$, yielding $\sigma_{tot}=74.8\ mb$
at $\sqrt s=1.8\ TeV.$

Earlier, in Ref.[3], we have suggested a model for diffraction
dissociation based on the dipole pomeron. The basic idea [4] behing
that model was that the function $g(s)$ in the $u-$matrix (2) is
proportional to the product of the quark number in hadrons $A$
and $B,$
$$g(s)\sim n_An_b.$$

In single diffraction, only quarks q
from one hadron should be counted,
consequently $\sigma_{SD}\sim \sqrt g.$ (Since  diffraction
dissociation makes only a fraction of the multiple production, this
will be taken into account by an adjustable parameter, to be
denoted by $c.$)

It follows from unitarity that the contribution from inelastic
processes is governed by the overlap function $G(\rho,s)$, which in
the $u-$matrix approach has the form [2]
$$G(\rho,s)={\Im u\over{\mid 1-iu\mid^2}},$$
or more explicitely - in case of the dipole pomeron (2):
$$G(\rho,s)={g(s)e^{-x}\over{\big(1+ge^{-x}\big)}^2},$$
where $x={\rho^2\over{4\alpha'(b+L)}}.$

Accordingly, the single diffraction dissociation cross section is
$$\sigma_{SD}(\rho,s)=\int\sigma_{SD}(\rho,s)d^2\rho=
c4\pi\alpha'(b+L){\sqrt g\over{(1+g)}}=cg^{-1/2}\sigma_{in}(s).$$
The free parameter $c$ can be eliminated by taking the ratio of the
cross sections at two differnt energies:
$$\sigma_{SD}(s_2)=\sigma_{SD}(s_1){\sigma_{in}(s_2)\over
{\sigma_{in}(s_1)}}\bigg({s_1\over{s_2}}\bigg)^{\epsilon/2}.$$

By using the values $\sigma_{SD}(\sqrt s=53\ GeV)=7.3\pm 0.4\ mb,
\ \sigma_{in}(\sqrt s=53 \ GeV)=31.29\ mb $ and $\sigma_{in}(\sqrt
s=546\ GeV )=47.68\ GeV, $ we find $\sigma_{SD}(\sqrt s=546\
GeV)=9.67\pm0.5\ mb. $

To conclude, the measurement of the single diffraction cross section
at Fermilab supports the concep of a moderate rise of the
hadronic cross sctions. The unitarized dipole pomeron model use  in
this paper is a typical representative of this class of models.

\bigskip
\bigskip
\centerline{\bf REFERENCES}
\hskip 2 cm
\item {1.} F.Abe et al. (The CDF
Collaboration), Measurement of $\bar pp$ Singe Diffraction
Dissociation at $\sqrt s=546$ and $1800$ GeV.  Fermilab-Pub-93/233-E,
CDF preprint.
\item{2.} L.L.Jenkovszky, B.V.Sruminsky and A.N.Vall,
Fiz.Elementarnyh Chast. i At.Yad. {\bf 19}(1988)180 (English transl.:
Sov.J.Particles \& Nuclei, {\bf 19} (1988)77).

\item{3.} L.L.Jenkvoszky and B.V.Struminsky, QCD and Diffraction (in Russian),
in: Proceedings of the VIII International Confernce of Quantum Field
Theory, Dubna 1987, p.160.
\item {4.} S.Barshay and H.Kondo,
Phys.Letters {\bf B177} (1986) 441.

 \end